# A novel framework for assessing metadata quality in epidemiological and public health research settings

Christiana McMahon, BSc[1], Spiros Denaxas, Ph.D. [1]
[1] Farr Institute of Health Informatics Research; Institute of Health Informatics, University College London, London, United Kingdom

**Abstract**

*Metadata are critical in epidemiological and public health research. However, a lack of biomedical metadata quality frameworks and limited awareness of the implications of poor quality metadata renders data analyses problematic. In this study, we created and evaluated a novel framework to assess metadata quality of epidemiological and public health research datasets. We performed a literature review and surveyed stakeholders to enhance our understanding of biomedical metadata quality assessment. The review identified 11 studies and nine quality dimensions; none of which were specifically aimed at biomedical metadata. 96 individuals completed the survey; of those who submitted data, most only assessed metadata quality sometimes, and eight did not at all. Our framework has four sections: a) general information; b) tools and technologies; c) usability; and d) management and curation. We evaluated the framework using three test cases and sought expert feedback. The framework can assess biomedical metadata quality systematically and robustly.*

**Introduction**

Metadata describe the data and the process through which they were collected. The use of metadata standards promotes a systematic approach to describing metadata elements and their structure. Metadata support researchers in making robust inferences and quickening the process of translating research findings into recommendations for change in policy and practice.

Metadata are handled by stakeholders in epidemiological and public health research across the stages of the research data lifecycle. For example, data sharing practices vary amongst researchers and access to good quality metadata inclusive of data models can help researchers to better understand the data[1, 2]. At the data integration and harmonization stages, metadata can provide contextual information which enables the semantic alignment of disparate biomedical datasets[3]. Having a robust and standardized approach to metadata markup can help better support the documenting of research studies[4]. The quality of metadata is critical in enabling discoverability, data sharing and reuse as they often provide the contextual information needed to understand how the data were, for example, collected and when.

The inherent complexity and heterogeneity of clinical information can lead to difficulties when trying to process these data. Identifying and characterizing certain research datasets such as linked clinical data as potential additional data sources, remains a challenging aspect of epidemiological and public health research[5]. Metadata, in many instances, are not subject to the same level of the scrutiny as the research data to which they are associated. Inconsistencies between steps taken to manage and curate research data can potentially reduce the extent to which these data may be combined and used as part of more complex research inquiries[6]. A potential consequence of this is reduced accuracy in meta-analyses as researchers are unable to analyze all potential datasets for inclusion in their study[7]. A lack of uptake of standards coupled with inconsistent quality of metadata can render analyses of epidemiological research data problematic[8]. Given that metadata are key to the research process, there is a need to develop and integrate metadata quality assessment into the daily routines of stakeholders in epidemiological and public health research.

The aim of this work was to review different methods of metadata quality assessment and improve understanding of current practices in epidemiological and public health research. This study has four objectives: a) describe the current state of the art in metadata quality assessment of epidemiological and public health research data; b) identify key relevant metadata quality assessment dimensions; c) create a novel framework for assessing metadata quality for epidemiology/public health research data; and d) evaluate framework using test cases and engaging with stakeholders.

## Methods

**Literature review of biomedical and computer science databases**

We performed a literature review in July 2014 using cross-disciplinary databases to identify literature focusing on the assessment of biomedical metadata quality. The aim was to establish the current state of art and identify which formalized techniques are available to guide the assessment. We used the following biomedical and computer science databases, ACM Digital Library, BioMed Central, CINAHL Plus, Cochrane Library, EMBASE, Lecture Notes in Computer Science, JSTOR, PubMed, SCOPUS and Web of Science. The range of databases chosen included some not specific to public health and epidemiology to help capture literature which may be categorized into other fields of research such as librarianship or archive management. We included Google and Google Scholar to identify gray literature and reports. We also used forward citation tracking[9] to systematically check through reference lists to help source other literature.

To be included in the review, the literature had to available in English and be accessible openly or through use of institutional login. The manuscript had to describe clearly the method used to assess metadata and ideally provide examples of where the method has been applied. The searches were not restricted in any way in terms of publication date, format or location. In the review we recorded: a) title; b) aim; c) conclusion; d) method of quality assessment – a brief description of the approach to quality evaluation; and e) identified metadata quality dimensions. We used the following search terms: 'epidemiology', 'metadata', 'metadata quality assessment', 'metadata quality dimensions', 'metadata quality evaluation', 'public health', 'public health and epidemiology', 'quality assessment', 'quality evaluation'.

**Online stakeholder survey**

We collected qualitative data on how metadata are currently handled across the research data lifecycle and by whom. We targeted the online survey at stakeholders using epidemiological and public health metadata. We invited stakeholders to share the perceived challenges associated with creating, using and assessing metadata and indicate which quality dimensions were of importance to biomedical metadata. We also wanted to identify how metadata are currently assessed in public health and epidemiological research settings.

The online survey ran for four months between November 2014 and February 2015 with a set of reminder emails sent two weeks before the survey was due to close. The survey had five sections: demographics, metadata, tools and technologies, metadata usability, and quality assessment. In the tools and technologies section, the list of metadata standards was partially based on a list from the Digital Curation Centre's website[10]. Within the context of epidemiological and public health research, participants were asked to: a) indicate the technologies, tools and frequency of usage of metadata, b) identify and rank by importance key metadata usability dimensions, c) identify metadata quality assessment dimensions, and d) describe the main barriers in assessing metadata quality. We chose the themes and questions of our survey based on a combination of the review findings and current understanding of the role of metadata in epidemiological and public health research.

The survey was designed and developed using Research Electronic Data Capture[11] (REDCap) version 5.7.5, a web-based data capture tool enabling development of survey instruments and collection of data in a secure environment. We included open-ended questions in the survey to facilitate the capture of qualitative data[12].

**Framework definition and evaluation**

To analyze the qualitative results, we adopted a hermeneutic approach; grounded theory was applied and the themes were collated inductively and iteratively. The initial high level metadata quality assessment framework consisted of four sections: a) general information; b) tools and technologies; c) usability; and d) management and curation.

We evaluated the framework in two steps. Firstly, by applying it to the metadata from three established cohort studies as test cases – the Millennium Cohort Study[13] (MCS), Midlife in the United States[14] (MIDUS) and Danish National Birth Cohort[15] (DNBC) studies. In doing so, we were able to test how fit for purpose the high level framework was and identify its strengths, weaknesses, and areas of improvement. We iteratively revised the created framework based on results from the evaluation process. Secondly, we engaged with stakeholders in epidemiological and public health research. This helped us to identify further areas of improvement and consider potential implications associated with implementing the framework into stakeholders' daily routines.

**Results**

**Literature review**

The literature review identified 11 studies[16-26] and nine dimensions of quality. These included accessibility, accuracy/correctness, completeness/totality, conformance, interoperability, consistency, frequency, timeliness, and meta-metadata. The most commonly occurring dimension was completeness/totality with all studies incorporating this dimension. The review also identified several studies[16, 21, 22, 25] that had divided their indicators of quality according to different criteria, such as structure[21]. In another study[23], a series of logic rules categories had been used. These were: a) rules of inclusion; b) rules of imposition; and c) rules of restriction. In two studies[19, 20], simple Dublin Core[27] elements had been incorporated into the method of quality assessment. We did not identify any studies specifically aimed for epidemiological and public health data.

**Online stakeholder survey**

96 individuals globally submitted data using the survey; most of whom indicated they were employed by a university and located in Europe. The most commonly indicated role in public health and epidemiology was "Data User". Most of the respondents used metadata in the analysis stage while the fewest respondents used metadata in the data destruction phase. Regarding the granularity of the metadata, 'Research study level' was most popular followed by 'Variable level'. The most commonly selected type of metadata was descriptive with a total of 37 votes followed by administrative with 30. Results also show, of those that submitted data, the most common format of metadata was PDF(s) with 27; the least commonly indicated was the Resource Description Framework[28] (RDF).

The perceived barriers to creating and/or using metadata can be categorized into the following: a) awareness of the issue of metadata quality and requiring the training and guidance to address these issues; b) inconsistencies in the formatting of metadata; c) inadequate tool availability which limited the ability to create research artefacts such as data dictionaries, and ability to extract metadata from older, previous versions of software; d) a lack of metadata quality standards; and e) inadequate resource availability, in terms of finances and time.

Of those who submitted data, most respondents, 46%, selected tools and technologies for use based on 'standard practice'. The clinical terminology most respondents had utilized was the International Classification of Diseases[29] followed by Medical Subject Headings[30]. The perceived difficulties when using clinical terminologies included: a) a lack of medical knowledge increasing the time taken to understand the terminology before use; and b) ease of use – the volume and complexity of clinical terminologies plus the potential for inaccurate clinical coding can sometimes render clinical terminologies difficult to use.

The most commonly used metadata standard by the respondents was the Data Documentation Initiative 2[31] (Codebook) followed by Data Documentation Initiative 3[32] (Lifecycle). A total of 11 people indicated that they used metadata catalogues to help improve the discoverability of their research. Of those who shared their experiences of using catalogues to improve discoverability, one of the challenges faced was the catalogue not being fit for use. The majority of those who submitted data, 20, viewed the metadata being available in an open access repository as being essential to usability whilst 17 people viewed being the metadata being standards-based as extremely important.

Perceived challenges respondents associated with assessing metadata quality in epidemiological and public health research can be categorized as follows. Firstly, there was felt to be a lack of guidance and awareness of metadata quality assessment. Several participants commented on the lack of guidance to assist quality assessment with one respondent commenting that one of the problems is identifying, "…the best way to determine quality" while another wrote, "I haven't thought about this". Secondly, there was a lack of knowledge - a total of three people commented on use of domain-specific terminology impacting how well they understood the metadata; for example, one respondent commented on a "lack of experience and knowledge of the area". Thirdly, respondents highlighted issues with knowing what software and other such tools were available to assist with quality assessments, and how to access these. Finally, the need to find the time needed to assess metadata quality was also identified[33].

'Accuracy' or the correctness of the metadata was deemed by respondents to be the most important metadata quality dimension. The least important quality dimension was 'Meta-metadata' with one respondent commenting, "before meta metadata let's get metadata clear and well understood" (Table 1).

**Table 1.** Dimensions of metadata quality for public health and epidemiological data as identified by the international stakeholder survey.

| Dimensions of metadata quality | Responses | | |
|---|---|---|---|
| | N | Percent | CI |
| Accuracy (correctness of the metadata) | 35 | 14.9% | 10-20 |
| Accessibility (extent to which the metadata can be accessed) | 34 | 14.5% | 10-20 |
| Discoverability (how visible the metadata are - can it be easily found) | 33 | 14.0% | 10-19 |
| Appropriateness (extent to which the metadata are relevant) | 24 | 10.2% | 7-15 |
| Comprehensiveness (extent to which the metadata are complete) | 24 | 10.2% | 7-15 |
| Interoperability (extent to which metadata can be exchanged and used without problems) | 21 | 8.9% | 6-13 |
| Timeliness (is the metadata current, inclusion of temporal information) | 20 | 8.5% | 5-13 |
| Versionability (extent to which a new version may be easily created) | 18 | 7.7% | 5-12 |
| Extendibility (extent to which the metadata may be easily extended) | 14 | 6.0% | 3-10 |
| Meta-metadata (metadata about the metadata) | 10 | 4.3% | 2-8 |
| Other | 2 | 0.9% | 0-3 |
| Total | 235 | 100.0% | |

Results of the survey show that most people, 19 out of the 37 who provided a response, sometimes assess the quality of the metadata they routinely handled; whilst, eight people never do so. Results also showed only one person out of the 28 who provided a response indicated they used some kind of metadata assessment criteria; this was a built-in diagnostics tool as part of a metadata editor.

**Framework definition**

We created a high-level metadata quality assessment framework for use in epidemiological and public health research settings.

The framework is composed of four sections: a) general information – to assess provision of the types, formats and granularity of the metadata available; b) tools and technologies – to assess the structure, application of clinical terminologies, indexing in catalogues, and use of Semantic Web technologies; c) usability – to assess presence in repositories, application of metadata standards, and provision of cross-walks or other semantic mappings; and d) management and curation- this refers to how the metadata were created and provision of and access to other metadata versions. Each section has three parts: 1) area of the framework; 2) associated quality dimensions; and 3) headings. Headings can have a basis in more than one quality dimension and the quality dimensions can be linked to more than one model (many-to-many relationships).

In the 'General information' section, completeness refers to how comprehensive the metadata are while granularity looks at the level at which the metadata are available, e.g. research study level. The types of metadata includes, administrative, descriptive, microdata and semantic. The formats include PDF, Spreadsheet, Word-processed document, XML, RDF and HTML. In having access to metadata in different formats, the discoverability and accessibility of research data may be potentially enhanced. Metadata provided in RDF for example, can be published on the World Wide Web and its inherent scalability can be harnessed facilitating the integration of multiple disparate resources in support of clinical research[34, 35].

In the 'Tools and technologies' section, accuracy refers to the use of clinical terminologies and in particular the provision of codes and category lists. Specifically, it suggests stakeholders look at the way in which the process of using clinical terminologies to encode clinical information is described. Structure refers to how the metadata are presented; for example, sweep-level metadata may be presented in tables with downloadable data dictionaries. Access to clearly structured and unambiguous metadata can help stakeholders to navigate through the metadata more easily. Accessibility can also be enhanced by including metadata in public facing, searchable catalogues such as the CALIBER portal[36]. Portals and catalogues can provide searchable, and where possible downloadable, records of metadata from different epidemiological and public health studies. The use of Semantic Web technologies such as biomedical ontologies can potentially enhance the interoperability, extendibility and discoverability of research data. Additionally, by using mechanisms such eXtensible Markup Language (XML), there is the potential to create newer versions of the metadata whilst referencing previous versions; an example of where this is possible is using the Data Documentation Initiative Lifecycle[32] metadata standard.

In the 'Usability' section, standards can refer to metadata standards such as Minimum Information for Biological and Biomedical Investigations[37] (MIBBI) or exchange standards such as Health Level Severn[38] (HL7). Cross walks refers to mappings between metadata standards such as Data Documentation Initiative 2[31] to Data Documentation Initiative 3[32]. In having cross walks between metadata standards, the metadata's interoperability could potentially be enhanced. Cross walks also refers to mappings between clinical terminologies such as MeSH and ICD. In providing these details, researchers wanting to run queries using multiple terminologies from disparate datasets could potentially identify which clinical codes are equivalent in support of clinical cohort phenotyping for example. The use of repositories acts as a mechanism to enhance discoverability and provide meta-metadata. Meta-metadata are metadata about the metadata; for example, date when metadata were created, the version number and by whom.

In the 'Management and curation' section, inclusion of the 'dates' and 'versions' headings encourage stakeholders to look at the meta-metadata and encourages them to assess how timely the metadata are. Provision of this meta-metadata also helps to support version control, and scope for extension. (Figure 1)

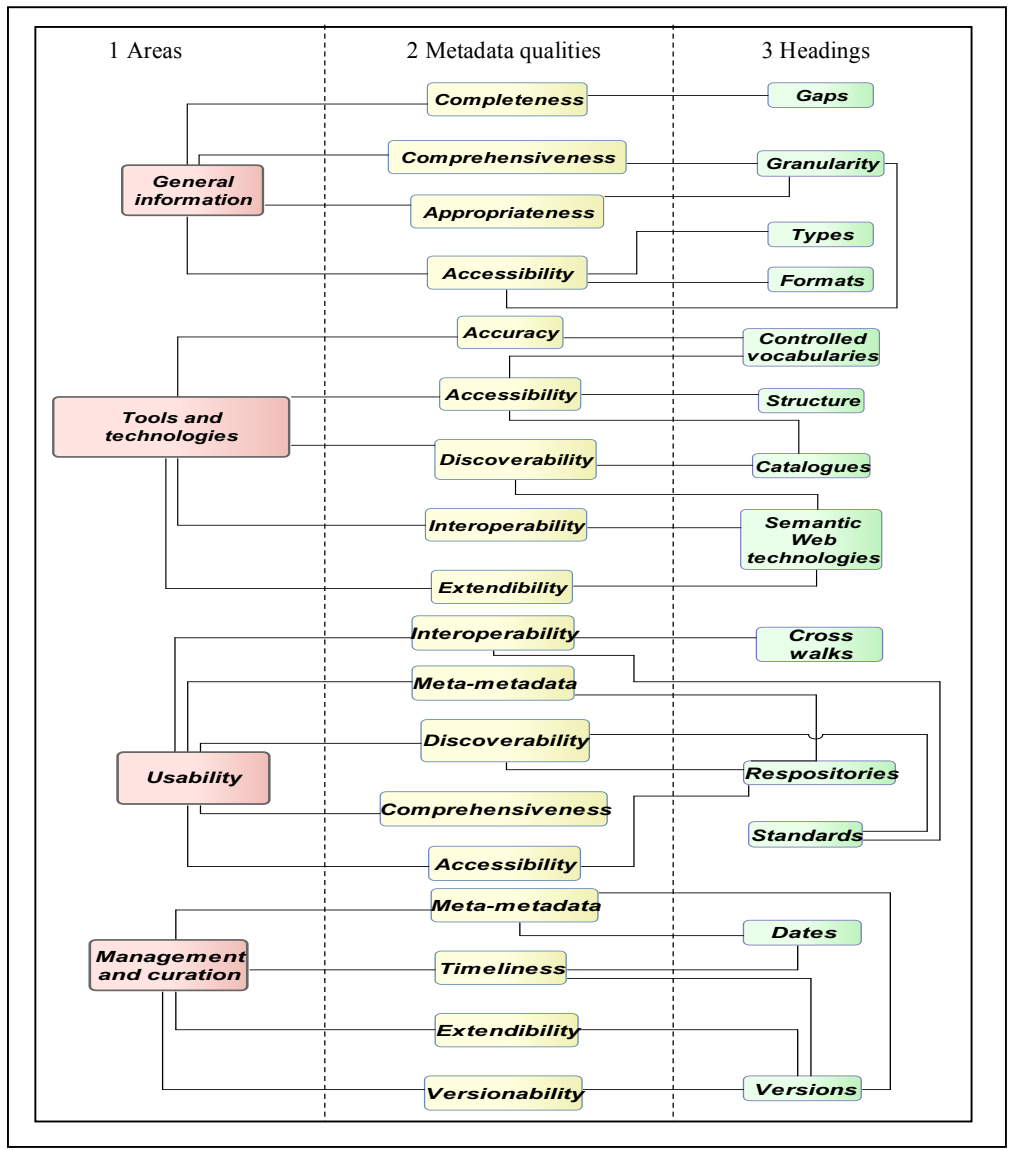

**Figure 1.** Sections of the high level quality assessment framework for metadata in epidemiological and public health research settings

**Framework evaluation**

We firstly evaluated the high level framework by applying it to three test cases: Millennium Cohort Study (MCS), Midlife in the United States (MIDUS), and the Danish National Birth Cohort (DNBC) study through an iterative process. After each iteration we revised the framework and re-applied it to the test-case studies.

When we applied the first version of the framework to the MCS study metadata; we noted that the 'General information' section did not include a mechanism to specify which artefacts were reviewed and where these can be found. To work around this, this information was placed under the 'Missing or incomplete metadata' section where the completeness of the metadata were reviewed. Though not ideal, we were able to record the artefacts reviewed and describe any difficulties experienced. In the second version, we added mechanisms to specify which documents were reviewed and in doing so, were able to record our findings more clearly. We also added sections for variable descriptions, online data visualization tools, links to other studies, sweeps or publications, and an 'other' section. By making these adjustments, we were able to record our findings more clearly. For example, we were able to record links to publications at sweep and study level for the MIDUS study, and links to all four sweeps, NCI indexed publications and a list of theses using DNBC data for the DNBC study.

In the 'Tools and technologies section of the first version of framework, we found that assessment findings could be recorded in different places due to the inherent links between headings and quality dimensions. For example, clinical terminologies can be structured using ontologies. Recording this information using the first version of the framework can use the 'Presence of clinical terminologies', 'Application of Semantic Web technologies', and 'Method of application and reason(s) for use' headings. We designed the framework to guide stakeholders when assessing metadata quality and not be a 'question and answer' style exercise; stakeholders are encouraged to record findings wherever they deem is most appropriate. In the second version, by adding the heading, 'Presence of code(s) and category(ies)' this enabled us to record the variable coding conventions for the MIDUS study, and the locations of the codebooks for interviews for the DNBC study. By merging the catalogues and repositories sections together, we were able to group this information together and record where metadata had been sourced.

In the 'Usability' section of the first version of the framework, we identified the need for a mechanism to record details of any caveats potentially impacting use of catalogues or repositories. Having access to this kind of information could help stakeholders better understand how research data could be accessed. We also identified the need to add a mechanism to record provision of the underlying metadata model; access to this kind of information could potentially enhance scope for metadata exchange – interoperability. We also recognized the potential for repetition in the first version of the framework by having separate 'repositories' and 'catalogue' headings. Hence in the second version, these headings were combined to reduce the potential for redundancies in the framework. In the second version we also added a mechanism to record metadata models. In making this change, we could record findings relating to how potentially interoperable the metadata are. For example, we could record that metadata for the MIDUS study were compliant with the, DDI, DC, MARC21 and Datacite metadata standards; metadata for the DNBC study were compliant with the DDI metadata standard.

Lastly, having applied the first version of the framework to the first test case, we decided that the 'Management and curation' section should be divided into two sections: a) the creation and version information of the metadata being reviewed; and b) the creation and version information of the review itself. In differentiating between these two types of meta-metadata, stakeholders can potentially see how up-to-date the metadata are, and how current the quality assessment is. Having implemented these changes, using the second version of the framework we were able to record the meta-metadata of the quality assessment itself and have a basic versioning system of the assessments using version numbers and dates. In having this meta-metadata, there is the potential to monitor metadata quality across assessments, and potentially identify any necessary corrective action.

**Stakeholder engagement**

We engaged with three stakeholders to obtain feedback and identify any further areas of refinement. It was noted by one stakeholder that the framework does not have a column for assessment findings. The framework should be viewed as a reference document and an altered version of the framework with a 'findings' column should be used when assessing metadata quality. The suggestion was also made to include potential answers for the headings; for example, for the heading, 'Indexing in catalogues/repositories' potential answers could include INDEPTH Data Repository iSHARE2[39] and IPUMS[40]. It was also unclear to a stakeholder which clinical terminologies were being referred to. The clinical terminologies heading was included in the framework to establish if clinical terminologies have been used and the version e.g. ICD 10. This could potentially support stakeholders in tracking any version changes e.g. ICD 9 to 10 and how these changes were managed. There is also the possibility to better support stakeholders in identifying any potential harmonisation work to enable continued use of these encoded research data. Therefore, we refined this area of the framework to better guide stakeholders when assessing metadata. Furthermore, a comment was made that finding metadata at a range of levels complete with links to publications is unrealistic and increased support for stakeholders is needed when assessing metadata. Whilst we acknowledge that the framework may provide a somewhat idealistic expectation of metadata, the framework was designed to provide a non-exhaustive list of headings to guide stakeholders. Context needs to be taken into consideration when assessing metadata quality as the granularity of metadata available may differ across the epidemiological and public health research domains.

**Discussion**

**Literature review**

In using a mix of computer science and biomedical databases we identified a range of literature for potential inclusion in the review. This was advantageous as in areas such as e-libraries and archive management, a lot of work had been done on metadata quality assessment and we were able to identify literature which, though not included in the review, provided useful contextual information. The literature review did not identify a method of metadata quality assessment designed specifically for use in an epidemiological and public health research setting.

However, given the methods identified were from research domains other than epidemiology and public health, the extent to which lessons learnt may be applied to epidemiological and public health research is inherently limited. This is because the methods identified lacked quality dimensions such as 'Extendibility'; potentially as this level of functionality was not required. Nevertheless, the review did identify quality dimensions, such as accuracy and accessibility, which at a high level are applicable to biomedical metadata. By combining quality dimensions such as these, with the dimensions we identified through the survey as important to biomedical metadata, such as 'Discoverability', we were able to produce a novel list of biomedical metadata quality dimensions.

**Online stakeholder survey**

Most of the respondents indicated they used descriptive metadata to some extent. This outcome could potentially reflect the respondents' roles in public health and epidemiology research and by extension their daily routines. Given that most of the respondents were data users employed by a university, and that metadata often accompanies epidemiological and public health research datasets, this could explain why descriptive metadata was the most popular. Results also showed that PDF was the most commonly handled format of metadata. This could be because organizations such as the Health & Social Care Information Centre provide the metadata for certain datasets as PDF documents. Given the proclivity of researchers to use these clinical data for research purposes, familiarity with this type of metadata is ameliorated.

The identified barriers to creating and/or using metadata in epidemiological and public health research could potentially be explained through a previous lack of focus on the importance of metadata; possibly contributing to the subsequent lack of training and support. Knowledge of the role of metadata across the stages of research data lifecycle is increasing as is familiarity with metadata standards. Stakeholders have a tendency to refer to best practice guidelines to inform the undertaking of research. This could explain why most respondents selected tools and technologies based on standard practice. This is important to the implementation and use of our novel framework in epidemiological and public health research as, integrating regular metadata quality assessments into work routines through inclusion in best practice guidelines, is a future goal of this work.

Regarding use of clinical terminologies, ICD was initially developed to encode clinical information for death certificates; whilst terminologies such as Read Codes and SNOMED CT were designed to encode information for clinical care. Both sources of data are routinely harnessed in public health and epidemiology research for secondary use but the current popularity of using ICD to encode clinical information could explain our survey findings. Of those who shared challenges experienced with using clinical terminologies, most respondents were data users, only nine were 'clinician/clinical advisor'. This could potentially explain why problems such as a lack of medical knowledge and ease of use were experienced. Having medical knowledge can potentially help when undertaking tasks such as clinical phenotyping, and often researchers work in multidisciplinary teams including clinicians to determine which clinical codes are needed.

Issues associated with the use of metadata catalogues could possibly be attributed to the lack of academic incentives to use this kind of catalogue coupled with limited knowledge of the existence of such platforms. Metadata catalogues can enhance the discoverability of research data and support stakeholders in identifying and characterizing data without accessing the research data directly. Of those who submitted additional aspects of importance for metadata usability, suggestions included having multiple formats of the metadata and semantic mappings. As metadata standards such as DDI are being continuously developed, semantic mappings or cross-walks between the different schemas are needed to support interoperability.

There are a number of strengths and weaknesses of the online survey. In using an online survey, we were able to contact and request participation from stakeholders globally within a relatively short period of time. We also requested the invitational email be forwarded to any interested parties potentially missed. However, this approach to recruitment rendered it infeasible to calculate a response rate and track to whom the survey was sent. There was also the potential for incorrect email addresses to limit the extent to which the invitational email was circulated.

Following the cross-disciplinary literature review and comprehensive online stakeholder survey, we identified a lack of formalized quality assessment methods for metadata in epidemiological and public health research settings. Through the literature review, we did not identify a framework specifically geared to epidemiological and public health research datasets and most approaches originated from information studies or other scientific domains. The stakeholder survey revealed that this lack of standardized methods for assessing metadata quality was acting as a barrier to assessing quality of metadata in epidemiological datasets in a systematic manner. Our novel framework is the first of its kind aimed specifically for use in epidemiological and public health research settings.

**Framework**

Our results show the five dimensions of most importance are: accuracy, accessibility, discoverability, appropriateness, and comprehensiveness.

To assess accuracy, stakeholders could establish how timely the metadata are and when the metadata were updated. To assess accessibility and discoverability, stakeholders could determine how well structured the metadata are and if they are available through a catalogue. Stakeholders are advised to establish whether encoding standards such as ICD and/or exchange standards such as HL7 have been utilized. It is also possible to assess how well knowledge has been managed through effective application of biomedical ontologies and whether these can be viewed. To assess appropriateness, stakeholders could determine whether necessary details have been provided to enable other stakeholders to understand and potentially reuse the research data. More specially, stakeholders are advised to determine how comprehensible the metadata are at different levels; for example, stakeholders could assess whether sufficient descriptions have been provided of variables to enable potential secondary researchers to characterize the dataset. Comprehensiveness could be determined by establishing how encompassing the metadata are and if the metadata contain gaps. For example, stakeholders could assess whether details of changes made to a sweep of data have been communicated effectively.

However, the definitions and recommendations presented are generic and potentially applicable across all aspects of public health and epidemiology research. For example, accuracy of metadata associated with electronic health records may not necessarily be the same as that of metadata associated with cohort studies; nor is the approach to quantifying accuracy of the metadata and the mechanism through which it may be tested. The difficulty remains in being able to deliver pragmatic quality dimensions whilst being mindful of the intricacies of the sub-domains of epidemiological and public health research.

**Framework evaluation**

Metadata standards such as DDI are commonly applied to metadata indexed in repositories/catalogues. Though this is beneficial as the metadata elements are standardised, and there is scope to download the DDI compliant XML, this does emphasize the need for increased application of metadata standards across epidemiological and public health research regardless of whether metadata are indexed. However, this in itself presents challenges such as having access to the necessary resources and assigning responsibility for ongoing maintenance.

Furthermore, for each test case, we were unable to find the method of application and reason(s) for use of Semantic Web Technologies; potentially as it does not seem commonplace to provide this kind of information. We had decided to include this section in the framework as we wanted to see if the method of application is linked in some way to the quality of the metadata. As stakeholders are able to customize the framework to suit local needs, we decided to keep this section as it is not a mandatory field in the assessment.

A potential weakness of the framework is that given multiple research artefacts may need to be sourced before quality can be assessed, the question of how much metadata is needed is raised. Another potential weakness of our framework is the lack of quantitative analysis; all assessments are manual/qualitative which are inherently subjective.

**Conclusions**

The use of clinical data for research purposes to better inform the development of clinical policy and practice is critical in public health and epidemiological research. One of the main challenges in assessing quality in epidemiological and public health research is a lack of awareness of the issue of poor quality metadata and the potential implications this can have on research data discoverability. Improved awareness of the issue of metadata quality is needed, as are mechanisms to integrate metadata quality assessments into daily routines of stakeholders in epidemiological and public health research.

We created and evaluated a novel framework as a platform-independent method of assessing metadata quality, with the goal of improving metadata in epidemiological and public health research settings and enhancing the potential for data discovery and reuse in the context of epidemiological and public health research studies.

Our next steps include engaging with stakeholders to establish a set of requirements for a series of computational metrics. Our short term goal is to identify a set of quantitative measures of quality to compliment the framework. The longer term goal is to use these metrics to increase objectivity and automate/quicken the overall assessment process.